\documentclass[aps,prx,twocolumn,groupedaddress,longbibliography]{revtex4-1}

\usepackage{graphicx}
\usepackage{bm}
\usepackage{amsmath}

\DeclareMathOperator{\Tr}{Tr}

\begin{document}

\title{Comparing skyrmions and merons in chiral liquid crystals and magnets}

\author{Ayhan Duzgun}
\author{Jonathan V. Selinger}
\affiliation{Department of Physics and Liquid Crystal Institute, Kent State University, Kent, Ohio 44242, USA}
\author{Avadh Saxena}
\affiliation{Theoretical Division, Los Alamos National Laboratory, Los Alamos, New Mexico 87545, USA}

\date{November 20, 2017}

\begin{abstract}
When chiral liquid crystals or magnets are subjected to applied fields or other anisotropic environments, the competition between favored twist and anisotropy leads to the formation of complex defect structures.  In some cases, the defects are skyrmions, which have $180^\circ$ double twist going outward from the center, and hence can pack together without singularities in the orientational order.  In other cases, the defects are merons, which have $90^\circ$ double twist going outward from the center; packing such merons requires singularities in the orientational order.  In the liquid crystal context, a lattice of merons is regarded as a blue phase.  Here, we perform theoretical and computational studies of skyrmions and merons in chiral liquid crystals and magnets.  Through these studies, we calculate the phase diagrams for liquid crystals and magnets in terms of dimensionless ratios of energetic parameters.  We also predict the range of metastability for liquid crystal skyrmions, and show that these skyrmions can move and interact as effective particles.  The results show how the properties of skyrmions and merons depend on the vector or tensor nature of the order parameter.
\end{abstract}

\maketitle

\section{Introduction}

When chiral liquid crystals are confined in an anisotropic environment, they experience geometric frustration:  The chirality favors a twist in the director field, but the anisotropy favors a director orientation that is incompatible with twist~\cite{Kamien2001,Oswald2005}.  Because of this frustration, these liquid crystals form complex topological defect structures, with regions of twist separating regions of the favored orientation.  In some cases, these defects are walls with a one-dimensional (1D) twist of the director field $\bm{n}(x)$~\cite{Smalyukh2005}.  In other cases the defects are skyrmions, which have a 2D variation of the director field $\bm{n}(x,y)$ with double twist going outward from the center, covering all possible orientations on the unit sphere~\cite{Bogdanov1998,Bogdanov2003,Fukuda2011,Ackerman2014,Leonov2014,Guo2016,Afghah2017,Ackerman2017b}.  In even more complex cases, the defects are hopfions, with a 3D variation of the director field $\bm{n}(x,y,z)$ in a knotted texture~\cite{Chen2013,Ackerman2017}.

An important feature of all three cases---walls, skyrmions, and hopfions---is that the orientation varies in a topological configuration that cannot anneal away, but the magnitude of the order parameter remains constant.  Unlike typical topological vortices, there is no singularity where the magnitude goes to zero (or otherwise changes away from its bulk value~\cite{Schopohl1987}).  These nonsingular defects were originally proposed in nuclear physics~\cite{Skyrme1962}, and they are now studied extensively in condensed matter, especially in chiral magnets~\cite{Roessler2006,Muehlbauer2009,Yu2010,Lin2013a,Lin2013,Nagaosa2013,Banerjee2014}, where they have potential applications in magnetic memory, logic, low power information technology devices, microwave detectors and oscillators~\cite{Finocchio2016} as well as topological spintronics~\cite{Fert2017}.

The nonsingular defect structure of skyrmions can be compared with blue phases in chiral liquid crystals~\cite{Grebel1983,Hornreich1990,Fukuda2009}.  A blue phase consists of a periodic array of double-twist tubes.  Each tube has a 2D variation of the director field going outward from the center, similar to a skyrmion except that it covers only half of the unit sphere.  In that sense, the double-twist tubes can be regarded as half-skyrmions, which are called merons~\cite{Lin2015}.  There is one crucial difference between blue phases and skyrmions:  In a blue phase, the double-twist tubes are separated by disclinations, which are singularities in the director field, where the magnitude of nematic order changes away from its bulk value.  Hence, an important issue in chiral liquid crystals is how to understand the crossover between skyrmions and blue phases.  Why would a chiral liquid crystal form singular or nonsingular defect structures?

A further theoretical issue is how to compare skyrmions and blue phases in liquid crystals with analogous structures in chiral magnets.  Both liquid crystals and magnets have orientational order parameters with magnitudes and directions.  They can both exhibit nonsingular defects (with constant magnitude of the order parameter), as well as singular defects (with the magnitude vanishing or otherwise changing away from its bulk value).  The main difference between these materials is the symmetry of the orientational order parameter:  liquid crystals have a \emph{tensor} order parameter, while magnets have a \emph{vector} order parameter.  How does this difference of symmetry affect the skyrmions or blue phases that form in the material?

The purpose of this paper is to address these issues through theoretical studies of chiral liquid crystals and magnets.  In Sec.~II, we consider a simple analytic model for chiral liquid crystals, and show that there are four characteristic energy scales:  the energy associated with the magnitude of nematic order, the chiral interaction, the anisotropy, and the temperature.  We derive a phase diagram in terms of three dimensionless ratios of these energies.  This phase diagram includes classical results for blue phases with no anisotropy, and extends the analysis to include anisotropy.  It shows that blue phases are stable when the energy associated with the magnitude of nematic order is relatively low.  Skyrmions are not stable structures in this phase diagram, but they can be metastable when that energy scale is high.

In Sec.~III, we present Monte Carlo and relaxational dynamic simulations of the model for chiral liquid crystals.  These numerical simulations confirm the phase diagram derived through simple analytic approximations.  They also show the formation of skyrmions as metastable defects, with length scales that can be understood analytically.

In Sec.~IV, we extend the simple analytic model to describe chiral magnets, which have a Dzyaloshinskii-Moriya (DM) interaction term in the free energy, arising from Dresselhaus spin-orbit coupling.  In this case, there are four characteristic energies:  the energy associated with the magnitude of magnetic order, the chiral DM interaction, the anisotropy, and the applied magnetic field, while the temperature scales out of the problem.  We derive a phase diagram in terms of three dimensionless ratios of these energies, and show that this phase diagram is quite similar to previous results from more detailed numerical calculations.  The results are generally similar to the liquid-crystal case, but with one important difference:  in magnets, skyrmions can be stabilized by the competition between anisotropy and applied magnetic field.  This competition is not available in liquid crystals because of the {\it tensor} nature of the order parameter.

In this article, we only consider the formation of skyrmions or blue phases driven by bulk chirality, known as Dresselhaus spin-orbit coupling in the magnetic case.  We should note briefly that modulated structures can also be driven by another mechanism for broken inversion symmetry.  In liquid crystals, that mechanism is called polarity.  Polarity is often induced by surfaces, and the phenomenon of surface-induced modulations has been studied for many years~\cite{Meyer1973,Lavrentovich1995}.  More recently, spontaneous bulk polarity has also been found in certain liquid crystals, and theoretical research has predicted that bulk polarity can induce blue phases~\cite{Alexander2007,Shamid2014}.  In magnets, the analogous mechanism for broken inversion symmetry at surfaces is called Rashba spin-orbit coupling, and it has also been shown to favor the formation of skyrmions~\cite{Rowland2016}.  Although we have only investigated the comparison between chiral (Dresselhaus) defects in liquid crystals and magnets, we expect that the same considerations will apply to polar (Rashba) defect structures.

\section{Theory of chiral liquid crystals}

\subsection{Model}

We begin with the theory of chiral liquid crystals, as usually studied in the context of blue phases.  A liquid crystal is described by a tensor order parameter $\bm{Q}(\bm{r})$, which is related to the scalar order $S(\bm{r})$ and the director field $\bm{n}(\bm{r})$ by $Q_{\alpha\beta}=S(\frac{3}{2}n_\alpha n_\beta -\frac{1}{2}\delta_{\alpha\beta})$.  In Landau-de Gennes theory, the free energy density can be expressed in terms of $\bm{Q}$ as
\begin{eqnarray}
\label{freeenergy}
F&=&\frac{1}{2}a\Tr\bm{Q}^2+\frac{1}{3}b\Tr\bm{Q}^3+\frac{1}{4}c\left(\Tr\bm{Q}^2\right)^2\\
&&+\frac{1}{2}L(\partial_{\gamma}Q_{\alpha\beta})(\partial_{\gamma}Q_{\alpha\beta})- 2Lq_0 \epsilon_{\alpha\beta\gamma}Q_{\alpha\delta} \partial_{\gamma}Q_{\beta\delta}.\nonumber
\end{eqnarray}
Here, the first three terms represent the free energy of a uniform system, expanded in powers of the tensor order parameter.  These terms favor certain eigenvalues of $\bm{Q}$, which correspond to a certain magnitude of uniaxial nematic order.  The quadratic coefficient $a$ is assumed to vary linearly with temperature, while $b$ and $c$ are assumed constant with respect to temperature.  The fourth and fifth terms represent the elastic free energy associated with variations of $\bm{Q}$ as a function of position.  The fourth term penalizes splay, twist, and bend deformations equally, with an elastic coefficient $L$.  The fifth term favors a chiral twist of the nematic order, with a characteristic inverse length $q_0$ arising from the molecular chirality.  We neglect other possible elastic terms that give different energy costs for splay, twist, and bend, such as $\frac{1}{2}L_2  (\partial_{\alpha}Q_{\alpha\gamma})(\partial_{\beta}Q_{\beta\gamma})$.

In the context of blue phases, following the work of Grebel et al.~\cite{Grebel1983}, researchers normally rescale parameters to simplify the theory.  To motivate this rescaling, it is convenient to consider the specific temperature at which $a=0$.  This temperature is below the first-order isotropic-nematic transition, which occurs at a positive value of $a$.  At this temperature, the first four terms in the free energy favor a nematic phase with order parameter $S\sim |b|/c$, the free energy density of the nematic relative to isotropic phase is $F\sim b^4 / c^3$, and the core radius of a disclination in nematic order is $\xi\sim (Lc/b^2)^{1/2}$.  Hence, at general temperature, we rescale the $Q$ tensor, the free energy density, and all lengths by those characteristic values.  In particular, we define the scaled free energy density as $\tilde{F}=F c^3 / b^4$.  The theory then depends only on two dimensionless ratios, which are normally written as
\begin{equation}
t=\frac{27 a c}{b^2},\qquad\kappa =\sqrt{\frac{108 c L q_0^2}{b^2}}.
\end{equation}

The parameter $t$ is a dimensionless temperature, which represents the temperature-dependent quadratic coefficient $a$ relative to $b$ and $c$.  The parameter $\kappa$ is a dimensionless chirality, which represents the natural twist $q_0$ relative to the disclination core radius $\xi$.  We can express the same comparison in terms of energies.  The free energy density associated with the favored chiral twist is $L S^2 q_0^2$, while the free energy density of a disclination core is $L S^2 \xi^{-2}$.  Hence, $\kappa^2$ can be interpreted as the energy scale of the favored chiral twist relative to the energy scale associated with changing the eigenvalues of $\bm{Q}$ inside a disclination core.  A liquid crystal material with low $\kappa$ is usually called ``low chirality,'' but it could equally well be called ``stiff nematic order.''  Likewise, a material with high $\kappa$ is usually called ``high chirality,'' but it could be called ``soft nematic order.''

In many experiments, a liquid crystal is placed in an anisotropic environment, which favors some alignment of nematic order with respect to a certain axis, which we can call the $z$ axis.  If the anisotropy favors alignment along the axis, it is called ``easy axis''; if it favors alignment perpendicular to the axis, it is called ``easy plane.''  There are two common mechanisms for anisotropy.  First, an electric field can be applied along the $z$ axis, leading to a dielectric anisotropy.  This field alignment can be represented by an additional term in the free energy of
\begin{equation}
\Delta F=-\Delta\epsilon E^2 Q_{zz},
\end{equation}
with $F_\text{total}=F+\Delta F$.  This term gives easy axis anisotropy if $\Delta\epsilon>0$ and easy plane anisotropy if $\Delta\epsilon<0$.  Following the same argument as above, we can rescale this term as $\Delta\tilde{F}=\Delta F c^3 / b^4$ to obtain the dimensionless anisotropy
\begin{equation}
\alpha=\frac{\Delta\epsilon E^2 c^2}{|b|^3}.
\end{equation}

A second mechanism for anisotropy is to put a liquid crystal in a narrow cell, of thickness $d$, between two surfaces with strong anchoring.  Homeotropic anchoring gives easy-axis anisotropy on the bulk liquid crystal, while planar anchoring gives easy-plane anisotropy.  To see the analogy between field-induced and surface-induced anisotropy, suppose the nematic order at the midplane is tilted at a small angle $\theta$ with respect to the $z$ axis.  For field-induced anisotropy, the extra free energy density (relative to an untilted state) is $\Delta\epsilon E^2 S \theta^2$.  For surface-induced anisotropy, the extra free energy density is $L S^2 \theta^2 / d^2$.  Hence, the effect of surface-induced anisotropy is similar to field-induced anisotropy with an effective $\Delta\epsilon E^2\sim L S / d^2$, and effective $\alpha\sim (Lc)/(d^2 b^2)$.  Of course, this analogy is an approximation for small tilt, and may break down when the tilt becomes larger.

Our goal is now to determine what modulated structures of the $\bm{Q}$ tensor minimize the free energy.  In particular, does the system form nonsingular defects, such as walls, skyrmions, and hopfions, with approximately constant eigenvalues of $\bm{Q}$?  Or does it form blue phases, with double-twist tubes (or merons) separated by disclinations in which the eigenvalues change away from their bulk values?  The results must be controlled by the three dimensionless parameters $t$, $\kappa$, and $\alpha$.

As a minimal model to address this question, we consider a 3D nematic order tensor that depends only on two spatial coordinates, $\bm{Q}(x,y)$, with no dependence on the third spatial coordinate $z$, under field-induced anisotropy.  This model can describe walls and skyrmions, although not hopfions.  Furthermore, it can describe a simple version of blue phases as vertical double-twist tubes (merons) separated by vertical disclinations, although it cannot describe the cubic structure of real 3D blue phases.

\subsection{Simple analytic calculations}

As a first step in analyzing this model, we make assumptions about $\bm{Q}(x,y)$ in each of the possible structures and calculate the free energies.  By comparing the free energies, we determine a phase diagram in terms of $t$, $\kappa$, and $\alpha$.  Of course, we recognize that these assumptions are very simple.  For that reason, in the following section we verify the results through Monte Carlo simulations of the model.

\subsubsection{Isotropic phase}

In the isotropic phase, the system is disordered with $\bm{Q}=0$ everywhere.  The scaled free energy density is $\tilde{F}_\text{iso}=0$, and the anisotropy contributes $\Delta\tilde{F}_\text{iso}=0$.  (In this analysis, we neglect any slight paranematic order that might be induced by the anisotropy.)

\subsubsection{Vertical nematic phase}

In the vertically aligned nematic phase, the director is $\hat{\bm{n}}=\hat{\bm{z}}$, and the order tensor is $Q_{\alpha\beta}=S(\frac{3}{2}n_\alpha n_\beta -\frac{1}{2}\delta_{\alpha\beta})$.  From Eq.~(\ref{freeenergy}), the free energy becomes $F=\frac{3}{4}a S^2+\frac{1}{4}b S^3+\frac{9}{16}c S^4$.
Minimizing with respect to the order parameter $S$ gives $S_{\text{vnem}}=(-b+\sqrt{b^2-24 a c})/(6 c)$, and hence the scaled free energy density is
\begin{equation}
\tilde{F}_{\text{vnem}}=-\frac{\left(3+\sqrt{9-8 t}\right)^2 \left(3+\sqrt{9-8 t}-4 t\right)}{93312}.
\end{equation}
The anisotropy makes an additional contribution of $\Delta\tilde{F}=-\alpha Q_{zz}=-\alpha S$, which implies
\begin{equation}
\Delta\tilde{F}_{\text{vnem}}=-\frac{\alpha\left(3+\sqrt{9-8 t}\right)}{18}.
\end{equation}

\subsubsection{Planar nematic phase}

In the horizontally aligned nematic phase, the director is $\hat{\bm{n}}=\hat{\bm{x}}$.  Most of the analysis is the same as for the vertically aligned nematic phase, with the same order parameter $S_{\text{pnem}}$ and the same scaled free energy density $\tilde{F}_{\text{pnem}}$.  However, the anisotropy now contributes $\Delta\tilde{F}=-\alpha Q_{zz}=+\frac{1}{2}\alpha S$, and hence
\begin{equation}
\Delta\tilde{F}_{\text{pnem}}=\frac{\alpha\left(3+\sqrt{9-8 t}\right)}{36}.
\end{equation}

\begin{figure}
\includegraphics[width=\columnwidth]{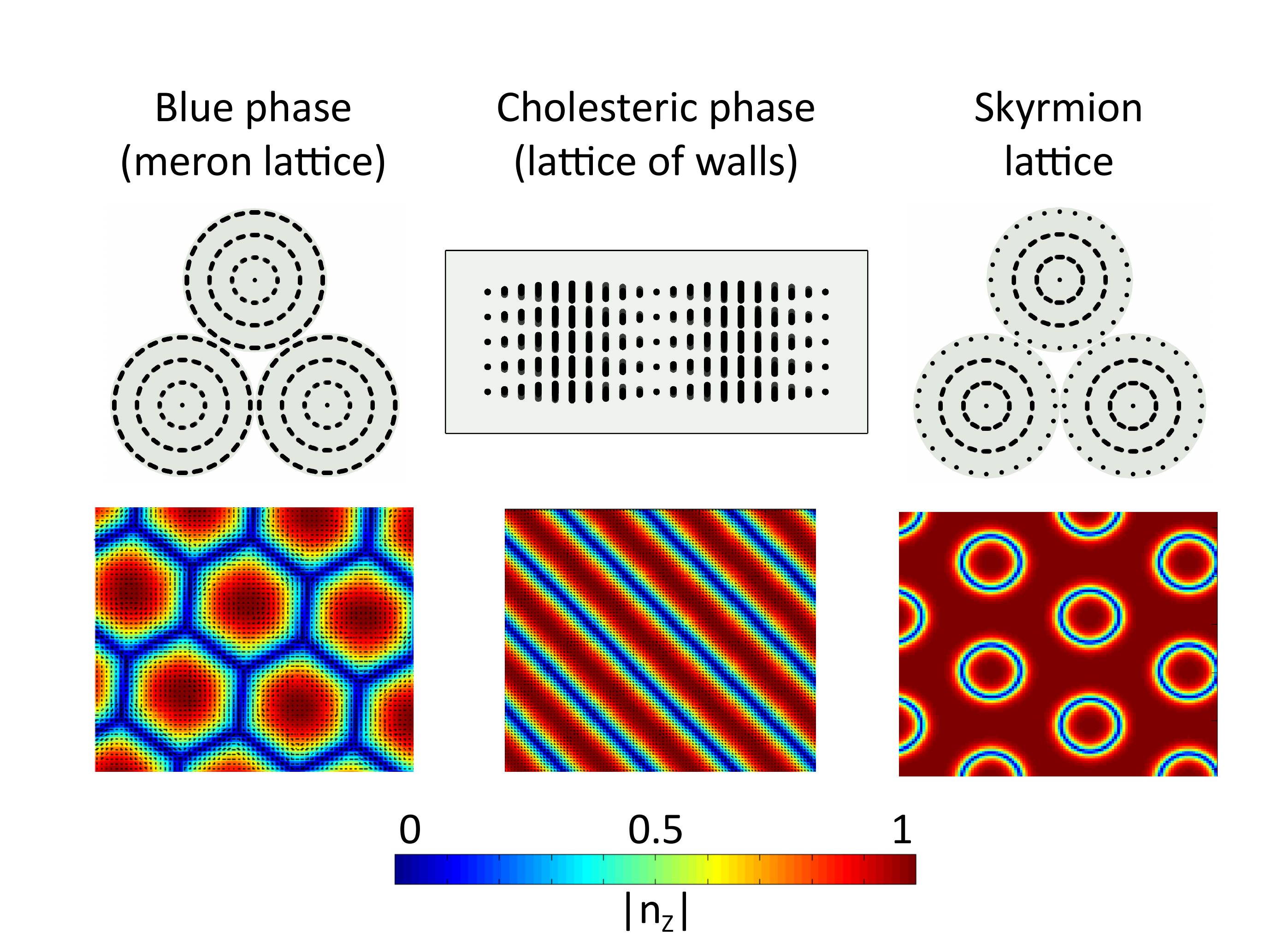}
\caption{(Color online) Structure of the modulated liquid-crystal phases studied in this paper:  blue phase (meron lattice), cholesteric phase (lattice of walls), and skyrmion lattice.  The top row shows schematic views of the director field, and the bottom row shows Monte Carlo simulation results (with the color scale indicating $|n_z|$).}
\label{modulatedphases}
\end{figure}

\subsubsection{Cholesteric phase (lattice of walls)}

A cholesteric phase has the twisted structure shown in Fig.~\ref{modulatedphases} (middle column).  It can be regarded as a periodic lattice of twist walls, separating regions in which the director field is aligned with the anisotropy.  To a first approximation, we assume that the director field is unperturbed by the anisotropy, so that $\hat{\bm{n}}(x)=-\hat{\bm{y}}\sin(\pi x/d)+\hat{\bm{z}}\cos(\pi x/d)$, where $d$ is the pitch.  The free energy is then 
$F=\frac{3}{4}a S^2+\frac{1}{4}b S^3+\frac{9}{16}c S^4+\frac{9}{4}\pi^2 L S^2 d^{-2}-\frac{9}{2}\pi L q_0 S^2 d^{-1}$.  By minimizing with respect to $d$ and $S$, we obtain $d_{\text{chol}}=\pi/q_0$ and $S_{\text{chol}}=(-b+\sqrt{b^2 -24 a c + 72 c L q_0^2})/(6 c)$, and the scaled free energy density becomes
\begin{align}
\tilde{F}_{\text{chol}}=-\frac{1}{93312}&\left(3+\sqrt{9-8t+6\kappa^2}\right)^2 \\
&\times\left(3-4t+3\kappa^2+\sqrt{9-8t+6\kappa^2}\right).\nonumber
\end{align}
The scaled free energy density associated with the anisotropy now depends on $x$, and it averages to $\Delta\tilde{F}=-\alpha \langle Q_{zz}\rangle=-\frac{1}{4}\alpha S$, giving
\begin{equation}
\Delta\tilde{F}_{\text{chol}}=-\frac{\alpha\left(3+\sqrt{9-8 t +6 \kappa ^2}\right)}{72}.
\end{equation}

\subsubsection{Blue phase (meron lattice)}

In 2D, a blue phase has the structure shown in Fig.~\ref{modulatedphases} (left column).  It consists of a hexagonal lattice of double-twist tubes, which can be regarded as merons or half-skyrmions.  In each meron, the director twists through an angle of $\pi/2$, from a vertical orientation at the center to a horizontal orientation at the edge of the tube.  A simple assumption for this variation can be expressed in cylindrical coordinates as $\hat{\bm{n}}(r)=-\hat{\bm{\phi}}\sin(\pi r/d)+\hat{\bm{z}}\cos(\pi r/d)$, for $0\le r\le d/2$, where $d$ is the diameter of the tube.  In each region between three tubes, the director field is in the $(x,y)$ plane, and it has a disclination with topological charge of $-1/2$.  The argument of Ref.~\cite{Schopohl1987} shows that the $\bm{Q}$ tensor becomes biaxial in the disclination core, but to a first approximation we will simply consider the core as an isotropic region.

To estimate the average free energy density of the blue phase, we represent each unit cell of the lattice (with area $A=\sqrt{3}d^2 /2$) as one meron (with $A=\pi d^2 /4$) and two disclinations (with the remaining area), and obtain
\begin{eqnarray}
\langle F\rangle&=&\frac{F_\text{meron}A_\text{meron}+2F_\text{defect}A_\text{defect}}{\sqrt{3}d^2 /2}\\
&=&\frac{\pi\sqrt{3}a S^2}{8}+\frac{\pi b S^3}{8\sqrt{3}}+\frac{3\pi\sqrt{3}c S^4}{32}\nonumber\\
&&+\frac{33.5 L S^2}{d^2}-\frac{3\sqrt{3} L q_0 S^2(4+\pi^2)}{4 d} \, . \nonumber
\end{eqnarray}
We then minimize over $d$ to find $d_\text{meron}=3.7/q_0$, and we use the same value of $S$ as in the cholesteric calculation.  The average scaled free energy density then becomes
\begin{align}
\langle F_{\text{meron}}\rangle=&-(9.7\times10^{-6})\left(3+\sqrt{9-8t+6\kappa^2}\right)^2 \times\nonumber\\
&\times\left(3-4t+4.1\kappa^2+\sqrt{9-8t+6\kappa^2}\right).
\end{align}
Similarly, the scaled free energy density associated with the anisotropy averages to
\begin{equation}
\langle\Delta F_{\text{meron}}\rangle=0.0027\alpha\left(3+\sqrt{9-8 t +6 \kappa ^2}\right) \, . 
\end{equation}

\subsubsection{Skyrmion lattice}

A hexagonal lattice of skyrmions is shown in Fig.~\ref{modulatedphases} (right column).  In each skyrmion, the director twists through an angle of $\pi$, from vertical at the center to horizontal and back to vertical at the edge.  A simple assumption for this variation can be expressed as $\hat{\bm{n}}(r)=-\hat{\bm{\phi}}\sin(2\pi r/d)+\hat{\bm{z}}\cos(2\pi r/d)$, for $0\le r\le d/2$.  In each region between three tubes, the director field is vertical, and hence there are no disclinations.

As in the previous case, we represent each unit cell of the lattice (with $A=\sqrt{3}d^2 /2$) as one skyrmion (with $A=\pi d^2 /4$) and two vertical nematic regions (with the remaining area), so that the average free energy density becomes
\begin{align}
&\langle F\rangle=\frac{F_\text{skyrmion}A_\text{skyrmion}+2F_\text{vnem}A_\text{vnem}}{\sqrt{3} d^2/2}\\
&=\frac{3 a S^2}{4}+\frac{b S^3}{4}+\frac{9 c S^4}{16}+\frac{100.4 L S^2}{d^2}-\frac{3 \sqrt{3} \pi ^2 L q_0 S^2}{2 d} \, . \nonumber
\end{align}
After the same minimization as in the previous case, we obtain $d_\text{skyrmion}=7.8/q_0$ and
\begin{align}
\tilde{F}_{\text{skyrmion}}=-&\frac{1}{93312}\left(3+\sqrt{9-8t+6\kappa^2}\right)^2 \times\\
&\times\left(3-4t+1.37\kappa^2+\sqrt{9-8t+6\kappa^2}\right).\nonumber
\end{align}
The anisotropy further contributes
\begin{equation}
\Delta\tilde{F}_{\text{skyrmion}}=-\frac{\alpha(8-\pi\sqrt{3})\left(3+\sqrt{9-8 t +6 \kappa ^2}\right)}{144}.
\end{equation}

\subsubsection*{Phase diagram}

We now have approximate algebraic expressions for the free energy $F_\text{total}=F+\Delta F$ for each of the six structures considered above, as functions of the three dimensionless variables:  temperature $t$, chirality $\kappa$, and anisotropy $\alpha$.  For each set of $(t,\kappa,\alpha)$, we determine which structure has the lowest free energy, and hence construct a phase diagram for the system.

\begin{figure}
\includegraphics[width=\columnwidth]{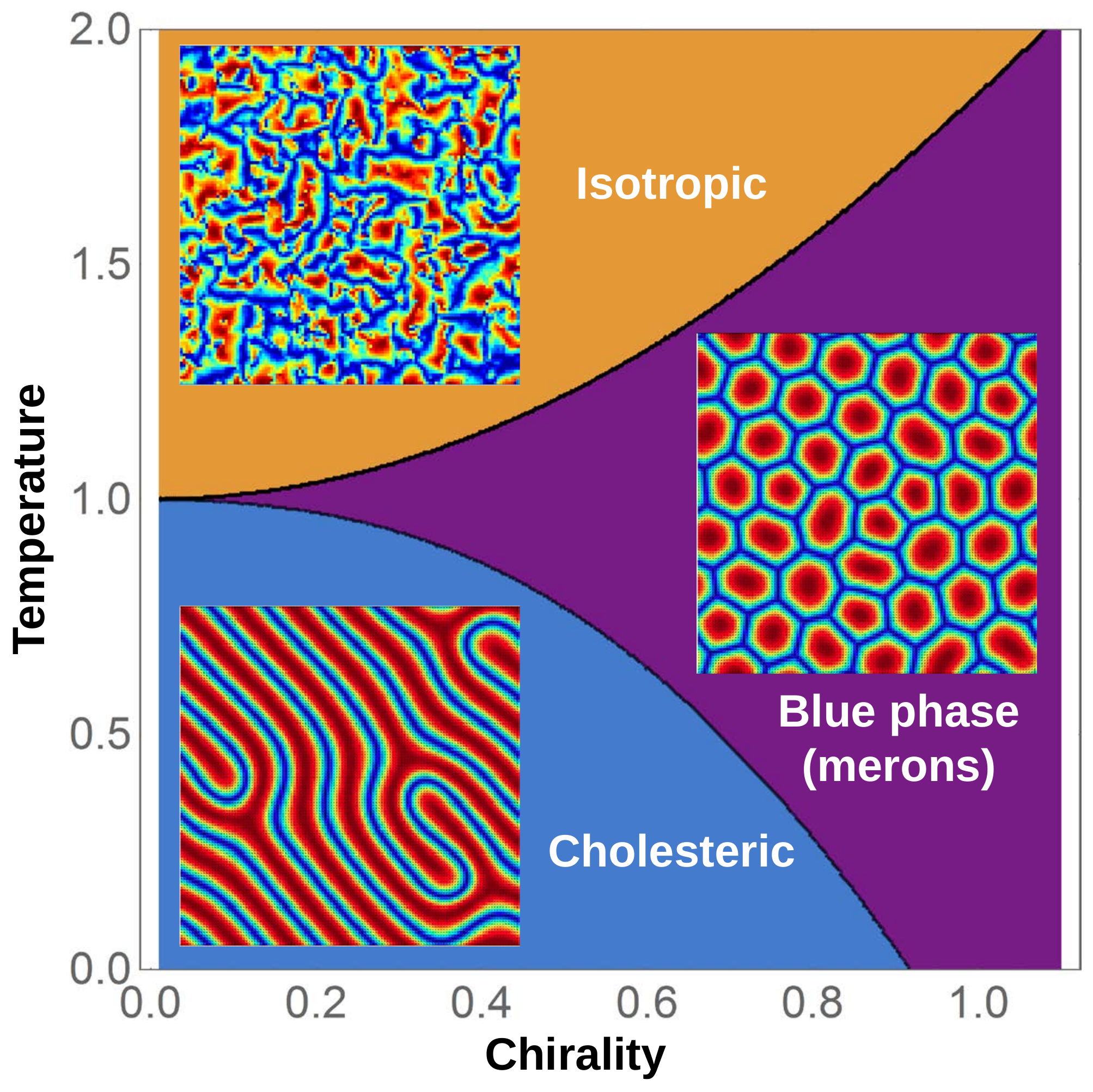}
\caption{(Color online) Phase diagram for chiral liquid crystals in the temperature-chirality plane, with no anisotropy.  The insets show structures calculated by the Monte Carlo 
simulations.  In those structures, the colors represent $|n_z|$, with the same color scale as in Fig.~\ref{modulatedphases}.}
\label{lcphasediagram2d}
\end{figure}

First, consider the case of no anisotropy, $\alpha=0$.  The phase diagram in the $(t,\kappa)$ plane is shown in Fig.~\ref{lcphasediagram2d}.  At high temperature, the system is in the disordered isotropic phase.  At lower temperature, for high chirality, the system forms a blue phase (meron lattice).  In this structure, there are favorable contributions to the free energy from the optimal magnitude of nematic order and from the optimal double twist of the director field within the merons.  There is an unfavorable contribution to the free energy from the disclinations between the merons, but these disclinations do not cost too much free energy because the nematic order is fairly soft in this case of high chirality.  By contrast, at low temperature and low chirality, the system forms a cholesteric phase.  In this structure, there are favorable contributions to the free energy from the optimal magnitude of nematic order and from the single twist of the director field (which is not as favorable as double twist).  There are no disclinations, which is reasonable because disclinations cost too much free energy when nematic order is stiff in this case of low chirality.

This phase diagram in the $(t,\kappa)$ plane is equivalent to the classic phase diagram for blue phases, which has been studied for many years.  In previous work, such as Ref.~\cite{Grebel1983}, it has been derived by methods that are much more rigorous than those in this section.  Here, we see that it is so robust that it occurs even with our very rough approximations.

\begin{figure*}
\includegraphics[width=2\columnwidth]{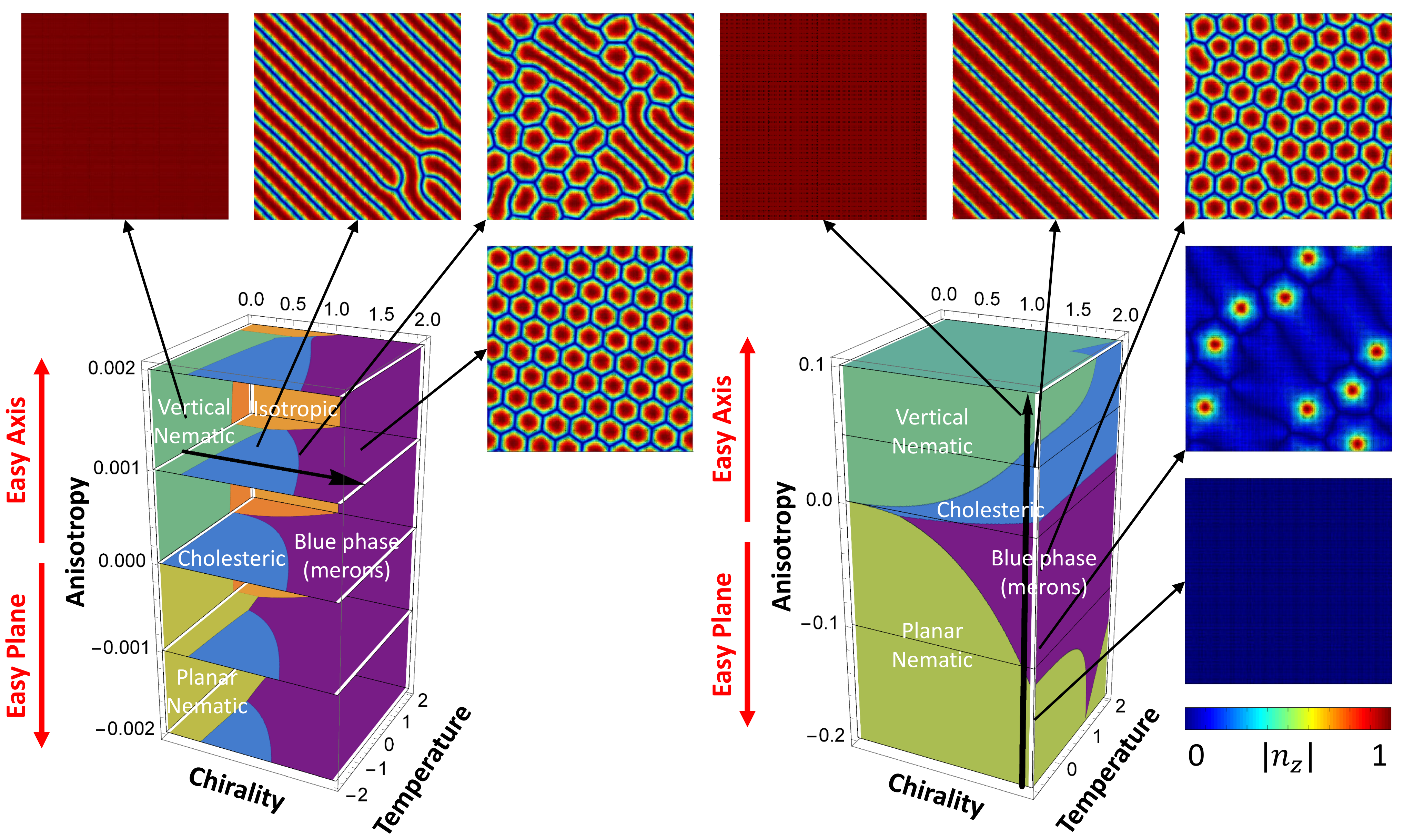}
\caption{(Color online) Two views of the phase diagram for chiral liquid crystals in the temperature-chirality-anisotropy space.  (Note that the scales on the axes are different in these two visualizations.)  The thick horizontal and vertical arrows show the Monte Carlo simulation paths discussed in Sec.~III, and the insets show structures calculated by the simulations.}
\label{lcphasediagram3d}
\end{figure*}

Now suppose that there is some anisotropy, which may be either easy-axis ($\alpha>0$) or easy-plane ($\alpha<0$).  We then obtain a 3D phase diagram in the $(t,\kappa,\alpha)$ space, which is shown using two different 3D visualizations in Fig.~\ref{lcphasediagram3d}.  (Note that the scales on the axes are different in these two views.)  This phase diagram still shows the isotropic, cholesteric, and blue phases.  In addition, the vertical nematic phase is stable for large $\alpha>0$, and the planar nematic phase is stable for large $\alpha<0$.  The transitions between uniform nematic phases and modulated phases (cholesteric or blue) depend mainly on the balance between anisotropy (which favors nematic) and chirality (which favors modulation).

The 3D phase diagram does not show any region in which the skyrmion lattice is stable.  At least with this set of rough approximations, the skyrmion lattice never provides the optimum balance among the different contributions to the free energy.  Even so, we can still ask:  Where in the phase diagram is the skyrmion lattice close to the optimum state?  That consideration will at least tell us when skyrmions are likely to be observed as metastable defects, and when they may even be stable if our approximations are not exactly correct (as with surface-induced anisotropy).  The answer is that the skyrmion lattice is almost the optimum state for very low chirality $\kappa$, near the transition between vertical nematic and cholesteric, which occurs for easy-axis anisotropy $\alpha>0$.  Hence, we can see that the skyrmion lattice and the meron lattice are actually very different types of structures, in spite of the fact that they look somewhat similar.  The meron lattice requires \emph{high} $\kappa$, so that the necessary disclinations will not cost too much free energy.  The skyrmion lattice requires \emph{low} $\kappa$, so that the nematic order parameter will be stiff against variations.

This analysis can be compared with a recent paper from our group~\cite{Afghah2017}, which modeled skyrmions using a very different theoretical formalism based on a director field (with constant order parameter $S$) in a 3D cell with strong homeotropic anchoring.  That paper found a phase diagram with three structures:  vertical nematic, cholesteric, and skyrmion lattice.  Although that phase diagram was expressed in terms of different variables, it can be translated into our current variables.  The nematic-cholesteric-skyrmion triple point in that phase diagram occurs at $(d/\xi)^2 \sim 10^3$ and $q_0 d \sim 10^{1/2}$, which implies $\alpha\sim(\xi/d)^2 \sim 10^{-3}$ and $\kappa\sim\xi q_0 \sim 10^{-1}$.  This result confirms that skyrmions form at low $\kappa$, in contrast with merons which form at high $\kappa$.

\section{Numerical Simulations}

\subsection{Equilibrium phases}

As an alternative method to minimize the free energy $F_\text{total}=F+\Delta F$, we run Monte Carlo (MC) simulations using the Metropolis algorithm.  In these simulations, the liquid crystal order is represented by a $3\times3$ traceless, symmetric tensor $\bm{Q}$ at each site of a 2D square lattice in the $(x,y)$ plane.  In the free energy, all derivatives are approximated by finite differences.  We relax the five independent components of $\bm{Q}$ by simulated annealing from a disordered state for each set of temperature $t$, chirality $\kappa$, and anisotropy $\alpha$.  The states found through this numerical method can then be compared with the states found by the simple analytic assumptions of Sec.~II(B).

As a first study, we vary the parameters $t$ and $\kappa$, for zero anisotropy $\alpha=0$, to explore the phase diagram of Fig.~\ref{lcphasediagram2d}.  At high $t$, the system is in the isotropic phase, with a highly disordered $\bm{Q}$ tensor field.  At low $t$ and low $\kappa$, the simulations show a cholesteric phase, with a lattice of twist walls separating vertically aligned stripes.  Because of fluctuations in the Monte Carlo simulation, the cholesteric order is not perfect, but rather exhibits hairpin defects.  At high $\kappa$, we find a blue phase, which consists of double-twist tubes or merons, separated by twist walls.  At each point where three walls intersect, there is a disclination of topological charge $-1/2$ in the orientational order.  These disclinations are points where the eigenvalue associated with the $\hat{\bm{z}}$ axis becomes dominant and negative in sign, surrounded by biaxial cores, as predicted in Ref.~\cite{Schopohl1987}).

For a second comparison, we vary $\kappa$ from 0.09 to 1.9, with the other two parameters fixed at $t=-$0.9 and $\alpha=0.001$.  This series of simulations moves along the thick arrow in Fig.~\ref{lcphasediagram3d} (left side).  A series of simulated structures is shown by the insets around the phase diagram.

At low chirality, the system is in the vertically aligned nematic phase.  The director is everywhere parallel to the anisotropy axis, $\hat{\bm{n}}=\hat{\bm{z}}$, as indicated by the uniform red color in the figure.  When the chirality increases, there is a transition into the cholesteric phase.  Because of fluctuations in the simulation, the cholesteric phase shows several dislocations in the stripe pattern, which correspond to disclinations in the orientational order.  As the chirality increases further, the density of dislocations increases, and the long stripes of vertical alignment evolve into shorter segments.  Eventually the segments shorten into hexagonal cells, which can be regarded as double-twist tubes or merons, separated by twist walls.  The transitions among these structures occur quite close to the phase boundaries predicted by the simple analytic assumptions.

For a third comparison, we vary $\alpha$ from -0.2 (easy plane anisotropy) to +0.1 (easy axis anisotropy), with the other parameters fixed at $\kappa=0.9$ and $t=-0.9$.  This series moves along the thick arrow in Fig.~\ref{lcphasediagram3d} (right side), with simulated structures shown by insets around the phase diagram.

For high easy plane anisotropy, the system is in a horizontally aligned nematic phase, with $\hat{\bm{n}}$ in the $(x,y)$ plane.  The orientation within the plane is random, and it is uniform across the system.  When the anisotropy is reduced toward zero, the system begins to show isolated merons, with vertical alignment in the center and double twist of the director going outward.  These merons are separated by large regions of $\hat{\bm{n}}$ in the $(x,y)$ plane, which must include disclinations in the planar director field.  As the anisotropy decreases further, the density of merons increases, and they eventually form a hexagonal lattice, which can be regarded as a blue phase.  After the anisotropy changes sign, and becomes larger in the easy axis direction, there is a transition into a cholesteric phase, with walls separating vertically aligned stripes.  For even larger easy axis anisotropy, the system forms a vertically aligned nematic phase, with a uniform director field.  Again, the transitions are generally consistent with the phase boundaries derived from the approximations of Sec.~II(B).

\subsection{Metastable skyrmions}

We do not see \emph{stable} skyrmions in the Monte Carlo simulations for any set of parameters in this model.  In that respect, the Monte Carlo simulations are once again consistent with the simple analytic calculations of Sec.~II(B):  One of the other phases is always lower in free energy than the skyrmion lattice.

Although skyrmions are not stable minimum energy states, they can still exist as \emph{metastable} states.  To investigate the possibility of metastable skyrmions, we run dynamic simulations of the same model with free energy $F_\text{total}=F+\Delta F$, with the code running on a graphical processing unit (GPU).  In these dynamic simulations, we integrate the $\bm{Q}$ tensor forward in time, following the relaxational equation $\partial Q_{\alpha\beta}(\bm{r},t)/\partial t=-\Gamma\delta F_\text{total}/\delta  Q_{\alpha\beta}(\bm{r},t)$, where $\Gamma$ is a mobility coefficient.  This equation is not required to conserve skyrmion charge, because the eigenvalues of $\bm{Q}$ can change in time.  However, it normally conserves skyrmion charge, unless the system goes over a significant energy barrier to changing the eigenvalues.

We begin the dynamic simulations with an initial condition corresponding to a circular skyrmion, in which the director is vertical at the center, and it twists by $180^{\circ}$ going outward to the perimeter.  Depending on the parameters relevant to  energetics, we see three types of shape evolution:  (a)~If the anisotropy is too large, the skyrmion shrinks and disappears; the final state is a vertical nematic.  (b)~If the anisotropy is too small, the skyrmion expands and evolves into one of the variations of cholesteric stripe patterns that was seen in the Monte Carlo simulations.  (c)~If the anisotropy is within the right range, the initial state relaxes into a metastable skyrmion.

\begin{figure}
\includegraphics[width=.9\columnwidth]{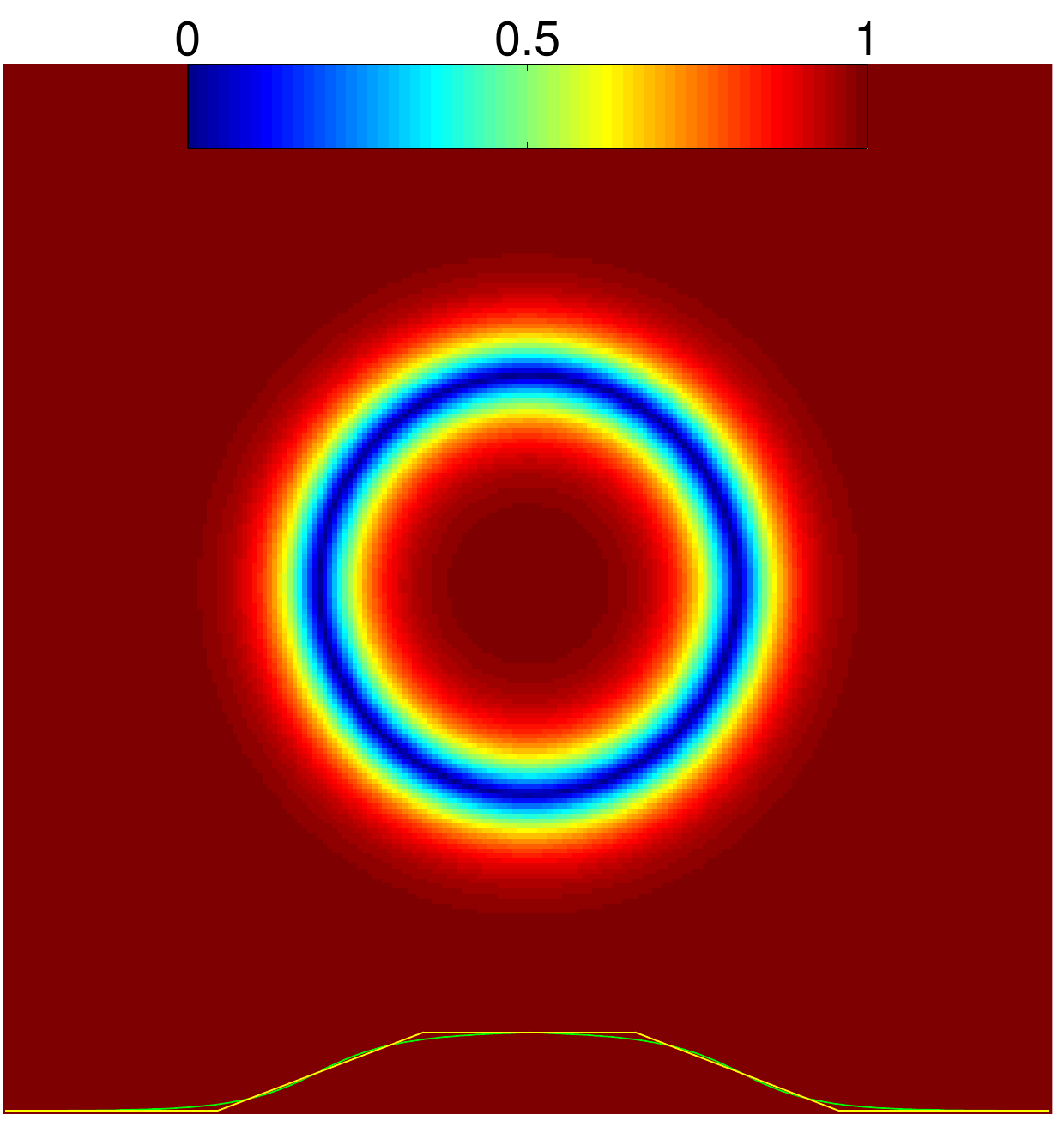}
\caption{(Color online) Simulation of a metastable skyrmion, with the color scale indicating $|n_z|$.  The yellow line on the bottom shows $n_z$ from $-1$ to $1$, as a function of $x$, for fixed $y$ in the center.  This structure can be regarded as a $\pi$-wall that is curved into a ring, with vertical nematic in the interior and the exterior.}
\label{metastable-skyrmion-simulation}
\end{figure}

The metastable skyrmion has the structure shown in Fig.~\ref{metastable-skyrmion-simulation}.  The director twists by $180^{\circ}$ from the center to the perimeter, but this twist is not uniform.  Rather, the director is almost vertical over some distance from the center and the twist occurs over a short range.  Hence, the skyrmion can be regarded as a $\pi$-wall that is curved into a ring, with a vertically aligned nematic phase in the interior and the exterior.

\begin{figure}
\includegraphics[width=.9\columnwidth]{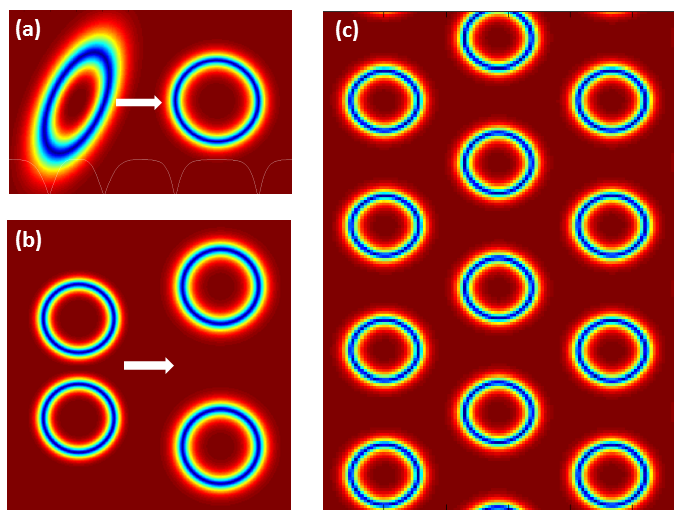}
\caption{(Color online) Static skyrmions as particles: (a)~An initially distorted shape quickly evolves into a circular ring. (b)~Skyrmions repel each other. (c)~A system of many skyrmions forms a lattice.}
\label{particle-like-figure}
\end{figure}

The size and shape of a skyrmion are very robust and long-lived. For example, in Fig.~\ref{particle-like-figure}, if the initial state is a distorted elliptical loop rather than a circle, the skyrmion quickly evolves into a final static circular shape which never breaks down.  If two skyrmions are in close proximity, they repel each other until they reach a separation comparable to the skyrmion diameter.  Because of this robustness and interaction, a system of many skyrmions forms a lattice, analogous to the crystallization of particles with repulsive interactions.  This behavior is similar to formation of a triangular or square lattice in simulations of magnetic skyrmions~\cite{Lin2013,Lin2015}.

To understand the metastable skyrmion structure, we represent the director field in cylindrical coordinates as $\hat{\bm{n}}(r)=-\hat{\bm{\phi}}\sin(\theta(r))+\hat{\bm{z}}\cos(\theta(r))$, and make the linear ansatz for the polar angle
\begin{equation}
\theta(r)=
\begin{cases}
0, & \text{for }r\leq r_\text{in},\\
(r-r_\text{in})\pi/\delta r, & \text{for }r_\text{in}\leq r\leq r_\text{out},\\
\pi, & \text{for }r\geq r_\text{out},\\
\end{cases}
\end{equation}
where $r_\text{in}$ is the inner radius of the ring, $r_\text{out}$ is the outer radius, and $\delta r=r_\text{out}-r_\text{in}$ is the thickness of the wall.  As in the calculations of Sec.~II(B), we calculate the free energy for this configuration using $Q_{ij}=S(\frac{3}{2}n_i n_j -\frac{1}{2}\delta _{ij})$, and we subtract the background energy of the vertical nematic phase.  We then make the substitution $g=r_\text{in}/\delta r$, to obtain a skyrmion free energy as a function of $g$ and $\delta r$.  Minimization with respect to $\delta r$ yields
\begin{equation}
\delta r=\frac{3\pi L q_0 S}{\Delta \epsilon E^2},
\end{equation}
showing that the wall thickness is determined by the competition between Frank elastic constant (which favors a thicker wall) and anisotropy (which favors a thinner wall).  The skyrmion free energy, relative to the vertical nematic, then becomes
\begin{align}
F=\frac{9\pi L S^2}{4}  &  \biggl[ \pi^2 \left(1-\frac{3L q_0^2 S}{\Delta\epsilon E^2}\right)(1+2g) +\log\left(1+\frac{1}{g}\right) \nonumber\\
& + \cos(2\pi g)\left[\text{Ci}(2\pi g)-\text{Ci}(2\pi (g+1))\right] \nonumber\\
& + \sin(2\pi g)\left[\text{Si}(2\pi g)-\text{Si}(2\pi (g+1))\right] \biggr], 
\label{fskyrmion}
\end{align}
where Ci and Si are the cosine integral and sine integral functions, respectively.

To minimize the skyrmion free energy over $g$, we rewrite the equation $\partial F/\partial g=0$ as
\begin{eqnarray}
\frac{3L q_0^2 S}{\Delta\epsilon E^2}&=&1-\frac{1}{\pi}\sin(2\pi g)\left[\text{Ci}(2\pi g)-\text{Ci}(2\pi (g+1))\right] \nonumber\\
&&+\frac{1}{\pi}\cos(2\pi g)\left[\text{Si}(2\pi g)-\text{Si}(2\pi (g+1))\right].
\end{eqnarray}
This equation has a solution provided that the ratio on the left side is between the minimum value
\begin{equation}
\lim_{g\to 0} \frac{3L q_0^2 S}{\Delta\epsilon E^2} =1-\frac{\text{Si}(2\pi)}{\pi}\approx0.55
\end{equation}
and the maximum value
\begin{equation}
\lim_{g\to\infty} \frac{3L q_0^2 S}{\Delta\epsilon E^2} =1.
\end{equation}
Equivalently, the range of anisotropy must be
\begin{align}
3 L q_0^2 S \leq \Delta \epsilon E^2 \leq 5.5 L q_0^2 S.
\end{align}
Within that range, skyrmions are metastable with a characteristic radius given by $r_\text{in}=g \delta r$.  For anisotropy below the lower limit of that range, the skyrmion radius will grow to infinity.  For anisotropy above the upper limit, a skyrmion will shrink and disappear.

\begin{figure}
\includegraphics[width=\columnwidth]{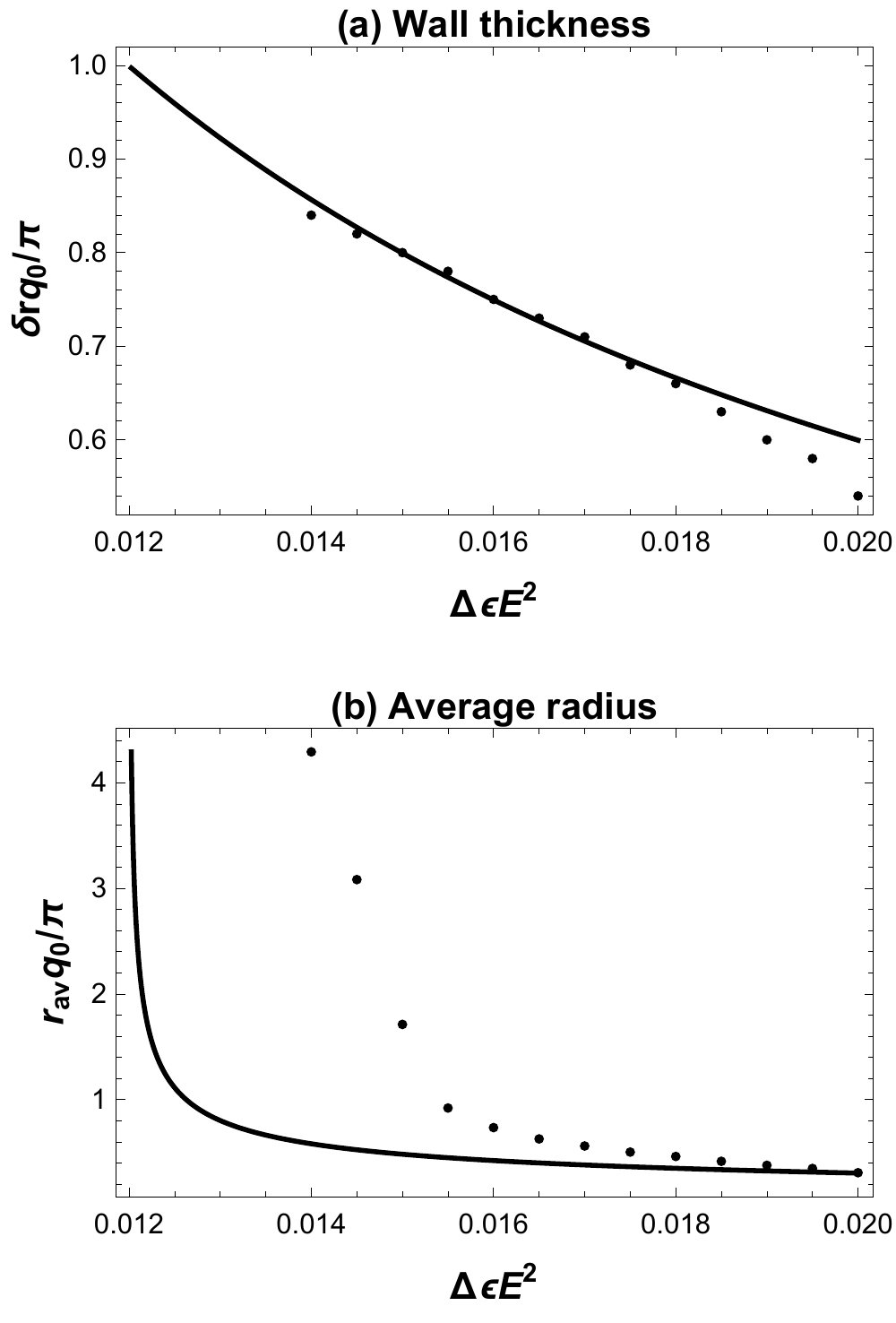}
\caption{Skyrmion wall thickness $\delta r$ and average radius $r_\text{av}=\frac{1}{2}(r_\text{in}+r_\text{out})$, as functions of the anisotropy $\Delta\epsilon E^2$, in units of $\pi/q0$.  The points represent simulation results, and the solid lines are the calculation in Sec.~III(B).  Parameters are $L=0.001$, $q_0 =\pi$, and $S=0.405$.}
\label{skyrmionradiusplots}
\end{figure}

This model for metastable skyrmions is qualitatively consistent with the simulations, which also find that metastable skyrmions can exist over a certain range of anisotropy.  As a further quantitative comparison, we plot the model calculations compared with simulation results for skyrmion wall thickness $\delta r$ and average radius $r_\text{av}=\frac{1}{2}(r_\text{in}+r_\text{out})$ as functions of the anisotropy $\Delta\epsilon E^2$ in Fig.~\ref{skyrmionradiusplots}.  The wall thickness calculations are in good agreement with simulation results over the full range of anisotropy that was simulated.  The average radius calculations are close to the simulation results for high anisotropy and small radius, where the simulated skyrmion is circular in shape.  However, for low anisotropy and large radius, there is a significant discrepancy; the model underestimates the minimum value of $\Delta\epsilon E^2$ for skyrmion stability.  This discrepancy seems to be caused by the shape of the skyrmions; the simulated skyrmion develops a four-fold anisotropy in this limit, perhaps because of the underlying lattice model.  Despite the latter discrepancy, the model generally provides a good estimate for the skyrmion size and the range of anisotropy needed for skyrmion stability.

\begin{figure}
\includegraphics[width=\columnwidth]{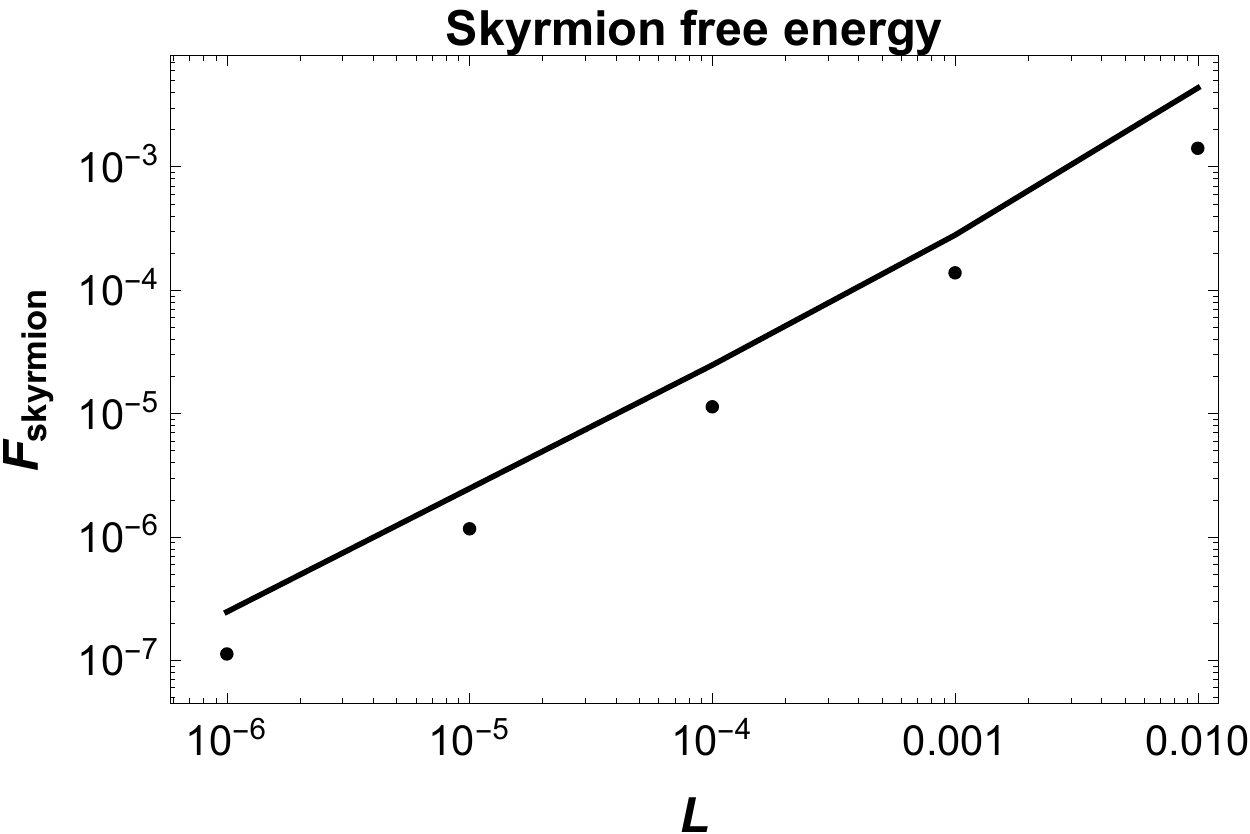}
\caption{Skyrmion free energy relative to the vertical nematic state, in arbitrary units.  The elastic constant $L$ is varied for fixed $a=-0.1$, $b=-3$, and $c=3$, with the anisotropy $\Delta\epsilon E^2$ adjusted to maintain the skyrmion size ($g$ and $\delta r$).  The points represent simulation results, and the the solid line is the calculation of Sec.~III(B) for the same $L$ and $\Delta\epsilon E^2$.}
\label{EnergyComp}
\end{figure}

For another comparison, we consider the free energy of a skyrmion, relative to the vertical nematic state.  This free energy difference is positive, indicating that skyrmions are metastable in this model.  To determine the magnitude of this difference in the simulation, we vary the elastic constant $L$ along the border of the vertical nematic and cholesteric phases.  For each $L$, we adjust the anisotropy $\Delta\epsilon E^2$ so that the size of the skyrmion is roughly the same.  In other words, the $g$ and $\delta r$ values are the same across simulations of different $L$ values.  The simulation results for free energy, relative to the vertical nematic, are shown by the points in Fig.~\ref{EnergyComp}.  By comparison, the calculation of Eq.~({\ref{fskyrmion}) for the same $L$ and $\Delta\epsilon E^2$ is shown by the solid line in the same figure.  These results are consistent up to a factor of 2, which is reasonable for such an approximate model.

As noted in the Introduction, many experiments have studied skyrmions in confined cholesteric liquid crystals~\cite{Bogdanov1998,Bogdanov2003,Fukuda2011,Ackerman2014,Leonov2014,Guo2016,Afghah2017,Ackerman2017b}.  These experiments generally cannot determine whether skyrmions are metastable, as predicted by the calculations in this section, or whether they are actually stable structures.  Indeed, that issue may depend on the exact form of the anisotropy, which can arise from an applied electric field or from homeotropic anchoring on surfaces.  Regardless of whether skyrmions are metastable or stable, they are separated from the uniform vertical state by a large energy barrier, and hence require significant disturbances in order to form or decay.  These skyrmions occur in liquid crystals with stiff nematic order, in contrast with blue phases (meron lattices), which occur in liquid crystals with soft nematic order or high chirality.

\section{Theory of chiral magnets}

In recent years, many investigators have carried out substantial theoretical research on modulated structures in chiral magnets, as in Refs.~\cite{Roessler2006,Muehlbauer2009,Yu2010,Lin2013,Nagaosa2013,Banerjee2014}.  In this section, we briefly review that work in a notation similar to the notation for chiral liquid crystals.  We then use this theory to compare magnetic skyrmions and merons with the analogous structures in liquid crystals.

A fundamental difference between liquid crystals and magnets is that liquid crystals are described by the tensor order parameter $Q_{\alpha\beta}(\bm{r})$, while magnets are described by the vector order parameter $M_\alpha(\bm{r})$, i.e. magnetization.  In Landau theory, the bulk free energy density of a chiral magnet can be written as
\begin{eqnarray}
F&=&\frac{1}{2}a|\bm{M}|^2 +\frac{1}{4}c|\bm{M}|^4 +\frac{1}{2}k(\partial_i M_j)(\partial_i M_j)\nonumber\\
&&+k q_0 \epsilon_{lik} M_l\partial_i M_k -H M_z -A M_z^2 .
\label{magneticfreeenergy}
\end{eqnarray}
Here, the first two terms represent the free energy of a uniform system, expanded in powers of the vector order parameter.  These terms favor a certain magnitude $|\bm{M}|$ of the magnetization.  The quadratic coefficient $a$ is assumed to vary linearly with temperature, while $c$ is assumed constant with respect to temperature.  The third and fourth terms represent the elastic free energy cost associated with variations of $\bm{M}$ as a function of position.  The third term penalizes all variations in $\bm{M}$, while the fourth term is a Dzyaloshinskii-Moriya interaction that favors certain twist deformations because of the Dresselhaus spin-orbit coupling.  The last two terms involve two distinct types of symmetry-breaking fields acting on the magnetic order.  The $H$ term is a standard magnetic field in the $z$ direction, which couples linearly to $\bm{M}$, while the $A$ term is a magnetocrystalline anisotropy, which couples quadratically to $\bm{M}$.  The anisotropy may be easy-axis with $A>0$, or easy-plane with $A<0$.

Equation~(\ref{magneticfreeenergy}) for the magnetic free energy is quite analogous to Eq.~(\ref{freeenergy}) for the liquid crystal free energy, but there are two important distinctions.  First, the bulk free energy for the liquid crystal has quadratic, cubic, and quartic terms, while the bulk free energy of the magnet has only quadratic and quartic terms.  Second, the liquid crystal has only a quadratic anisotropy acting on the order parameter, while the magnet has both a linear field and a quadratic anisotropy.  Both of these distinctions arise from the tensor vs.\ vector nature of the order parameter.

By analogy with the liquid crystal theory, we can simplify the magnetic theory by rescaling parameters.  Here, the characteristic value of the magnetic order parameter is $M\sim(|a|/c)^{1/2}$, the free energy density of the ferromagnetic relative to disordered phase is $F\sim a^2 /c$, and the core radius of a vortex in magnetic order is $\xi\sim(k/|a|)^{1/2}$.  Hence, we rescale $\bm{M}$, $F$, and all lengths by those characteristic values.  The theory then depends on three dimensionless ratios, which we write as
\begin{equation}
\kappa=q_0\sqrt{\frac{k}{|a|}},\qquad h=H\sqrt{\frac{c}{a^3}},\qquad\alpha=\frac{A}{|a|}.
\end{equation}
As in the liquid crystal case, the parameter $\kappa$ is a dimensionless chirality, which represents the natural twist $q_0$ relative to the disclination core radius $\xi$.  Equivalently, $\kappa^2$ can be interpreted as the energy scale of the favored chiral twist relative to the energy scale associated with changing the magnitude of $\bm{M}$ inside a defect core.  Low $\kappa$ can be called ``low chirality'' or ``stiff magnetic order,'' while high $\kappa$ can be called ``high chirality'' or ``soft magnetic order.''  The parameters $h$ and $\alpha$ are dimensionless versions of the field and anisotropy.  The anisotropy $\alpha$ is analogous to the anisotropy in the liquid crystal case, while the field $h$ does not exist in the liquid crystal.

The magnetic system does not have a temperature parameter $t$ analogous to the liquid crystal case.  Because the magnetic free energy density includes the quadratic and quartic but not the cubic terms in $\bm{M}$, the temperature scales out of the magnetic case, leaving a problem with no explicit dependence on the temperature-dependent coefficient $a$ (assuming that $a<0$ so that the system is in an ordered phase).

Many investigators have already studied the phases of this model through detailed numerical simulations.  We suggest that key features of the phase diagram can be understood through simple analytic calculations, analogous to the liquid crystal calculations in Sec.~II(B).  Hence, we repeat those calculations for the magnetic case, and compare the results with simulations from the literature.

For these simple analytic calculations, we consider the following phases:

\paragraph{Vertical ferromagnetic phase}
The system has uniform magnetic order in the $z$ direction, with $\mathbf{M}=M\hat{\bm{z}}$.  After minimizing over $M$, the scaled free energy density is $\tilde{F}_\text{vert}=-\frac{1}{4}-h-\alpha$.

\paragraph{Tilted ferromagnetic phase}
The magnetic order is given by $\textbf{M}=M[\hat{\bm{x}}\sin\theta+\hat{\bm{z}}\cos\theta]$, and the scaled free energy density is $\tilde{F}_\text{tilt}=-\frac{1}{4}-h\cos\theta-\alpha\cos^2 \theta$.  The tilt $\theta$ is determined by the competition between $h$ and $\alpha$.

\begin{figure}
\includegraphics[width=\columnwidth]{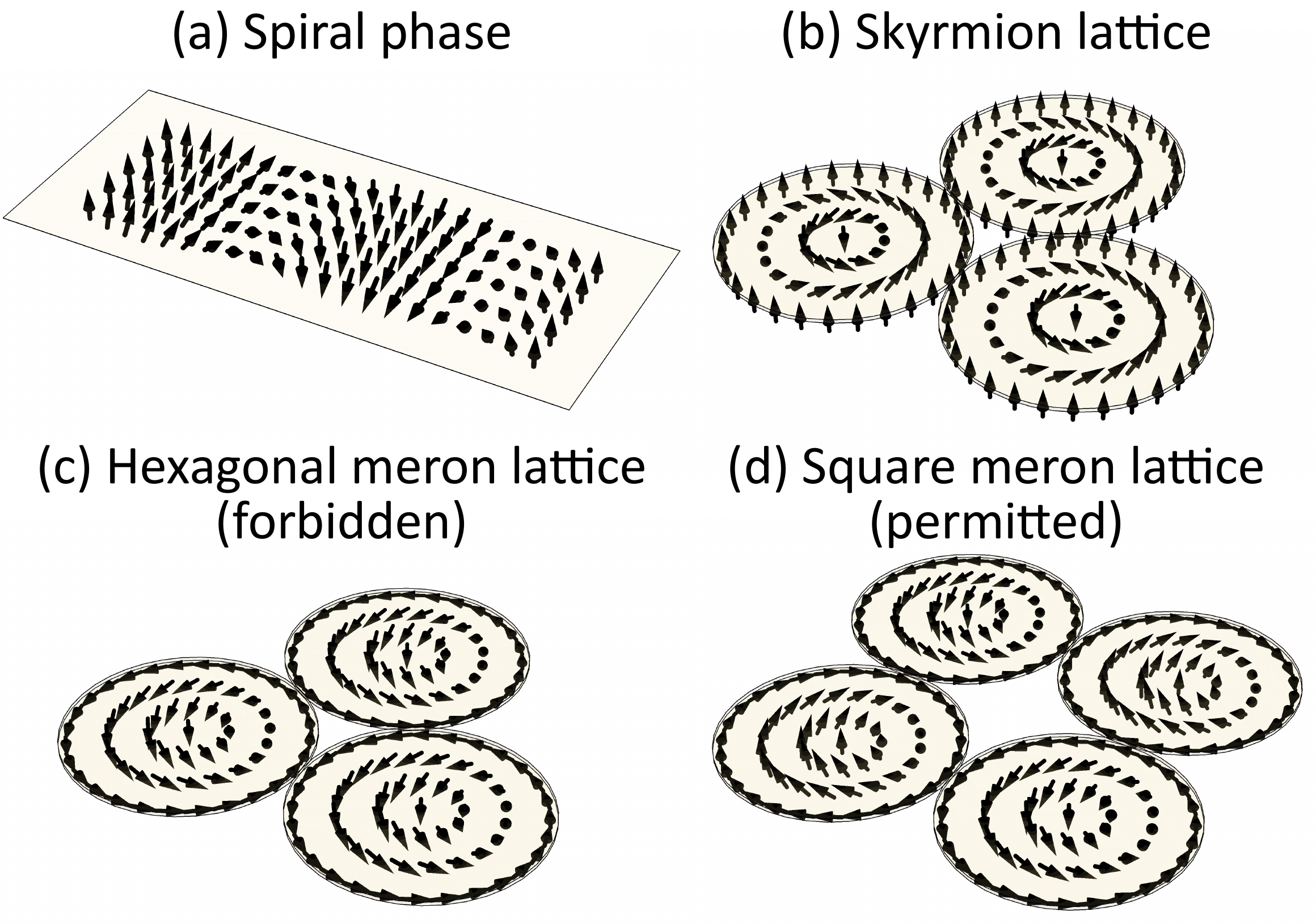}
\caption{Structure of the different modulated magnetic phases studied here.}
\label{magnetic_skyrmion_figures}
\end{figure}

\paragraph{Spiral phase}
The spiral phase of magnetic systems has the structure shown in Fig.~\ref{magnetic_skyrmion_figures}(a), analogous to the cholesteric phase of liquid crystals.  If $h$ and $\alpha$ are zero, the modulated structure is $\bm{M}(x)=M[-\hat{\bm{x}}\sin(\pi x/d)+\hat{\bm{z}}\cos(\pi x/d)]$.  If $h$ and $\alpha$ are small but nonzero, the structure is only slightly distorted, so that the expression can still be used as a first approximation.  After minimizing over $M$ and $d$, the average scaled free energy density is $\tilde{F}_\text{spiral}=-\frac{1}{4}(1+\kappa^2)^2 -\frac{1}{2}\alpha(1+\kappa^2)$.

\paragraph{Skyrmion lattice}
Skyrmions are modeled by disks arranged in a hexagonal lattice, as in Fig.~\ref{magnetic_skyrmion_figures}(b).  Within each disk, the magnetic order twists through an angle of $\pi$, from downward at the center to upward at the edge.  Our linear assumption for this variation is $\bm{M}(r)=M[-\hat{\bm{\phi}}\sin(2\pi r/d)+\hat{\bm{z}}\cos(2\pi r/d)]$, for $0\le r\le d/2$.  Between the disks, the magnetic order is uniform and upward.  After minimizing over $M$ and $d$, the average scaled free energy density is $\tilde{F}_\text{skyrm}=-\frac{1}{4} -0.36\kappa^2 -0.15\kappa^4 -0.093h -0.37h(1+0.80\kappa^2)^{1/2} -0.55\alpha -0.36\alpha\kappa^2$.

\paragraph{Meron lattice}
Merons are modeled by disks with a twist of $\pi/2$ from the center to the edge.  These disks \emph{cannot} be arranged in a hexagonal lattice, as shown in Fig.~\ref{magnetic_skyrmion_figures}(c), because the magnetic order parameter would be incompatible at each point where two disks meet.  In that respect, the vector order parameter for a magnet is quite different from the tensor order parameter for a liquid crystal.  As an alternative, merons \emph{can} be arranged in a square lattice, shown in Fig.~\ref{magnetic_skyrmion_figures}(d).  In this structure, there is a regular alternation of merons with the central $\bm{M}$ pointing upward or downward.  Our linear assumption for the variation within each disk is $\bm{M}(r)=M[-\hat{\bm{\phi}}\sin(\pi r/d)+\hat{\bm{z}}\cos(\pi r/d)]$, for $0\le r\le d/2$.  In each region between four disks, the magnetic order has a vortex of topological charge $-1$, which we model as a disordered, isotropic region.  After minimizing over $M$ and $d$, the average scaled free energy density is $\tilde{F}_\text{meron}=-\frac{1}{4} -0.36\kappa^2 -0.11\kappa^4 -0.23\alpha -0.28\alpha\kappa^2$.

\begin{figure}
\includegraphics[width=\columnwidth]{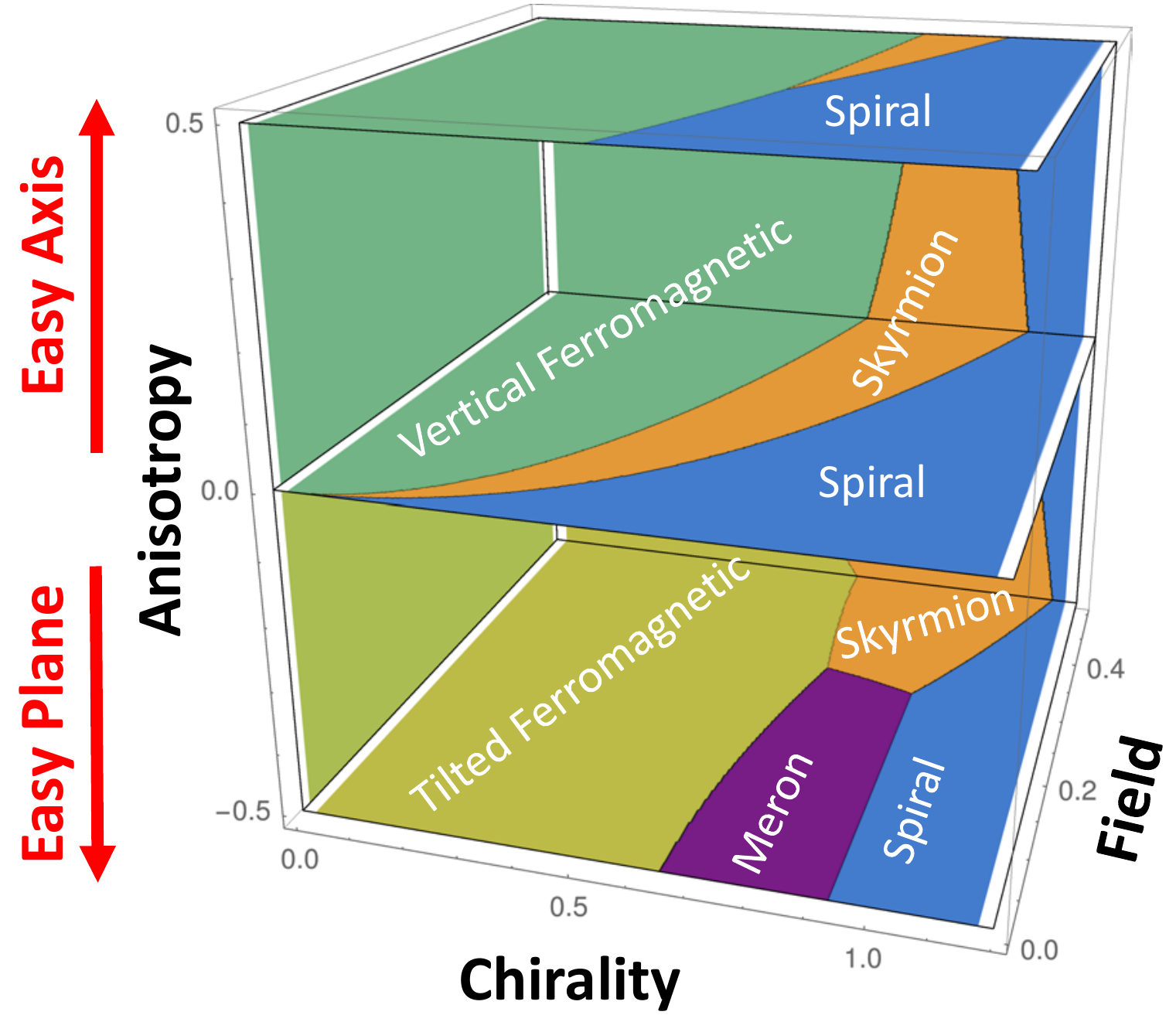}
\caption{(Color online) Visualization of the phase diagram for chiral magnets in the chirality-field-anisotropy space.}
\label{magneticphasediagram3d}
\end{figure}

By comparing the free energies of these structures, we construct a 3D phase diagram in the chirality-field-anisotropy space, as shown in Fig.~\ref{magneticphasediagram3d}.  In the limit of low chirality, the system forms a ferromagnetic phase, which is vertical for large easy-axis anisotropy and tilted for large easy-plane anisotropy.  In the limit of high chirality, the system forms a spiral phase.  The more complex skyrmion and meron lattices occur for intermediate chirality.  In this intermediate regime, easy-plane anisotropy favors the meron lattice, because this lattice has large planar regions.  A field favors the skyrmion lattice, because it has a predominant vector orientation which can align with the field.  

\begin{figure}
\includegraphics[width=\columnwidth]{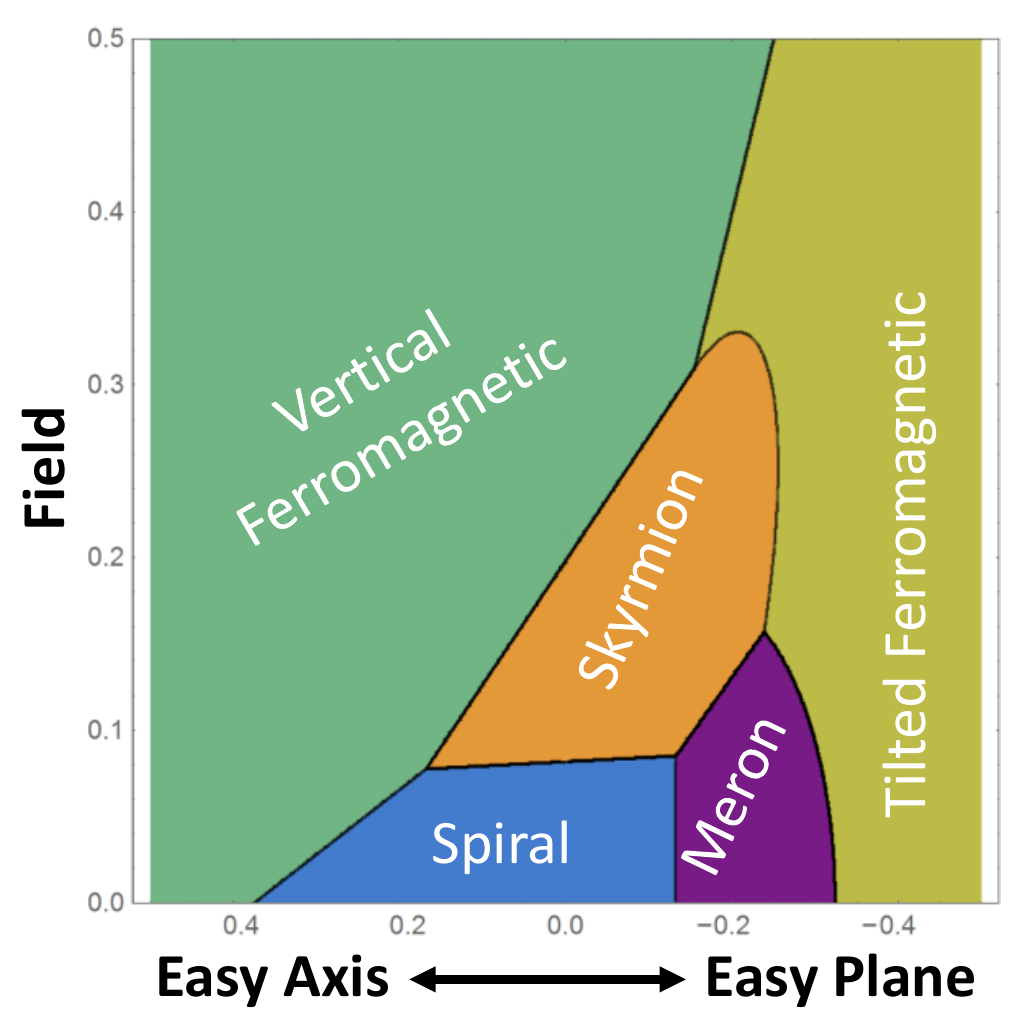}
\caption{(Color online) Cross section of the phase diagram for chiral magnets in the field-anisotropy plane for fixed chirality $\kappa=0.5$.}
\label{magneticphasediagram2d}
\end{figure}

Instead of performing our own simulations, we can compare the results of these approximate analytic arguments with previously published simulations by other investigators.  As examples, Refs.~\cite{Lin2015} and \cite{Rowland2016} both present phase diagrams for magnetic structures in the field-anisotropy plane, which can be compared with a cross section of our phase diagram for fixed chirality $\kappa=0.5$, as shown in Fig.~\ref{magneticphasediagram2d}.  We can see the same general arrangement of the phases in Fig.~7 of Ref.~\cite{Lin2015}, and in Fig.~1 (left) of Ref.~\cite{Rowland2016}, as in our Fig.~\ref{magneticphasediagram2d}.  (As a matter of terminology, the polarized ferromagnetic phase is equivalent to what we have called vertical ferromagnetic, and canted ferromagnetic is equivalent to what we have called tilted.)  Hence, we can observe that the approximate analytic arguments of this section capture key features of the free energy balance among the phases, even without the need to do detailed numerical simulations.

\section{Discussion}

The work presented in this article attempts to put the topological phases in liquid crystals and chiral magnets on the same footing.  It enables us to compare skyrmions with merons, and also to compare various orientational phases of chiral liquid crystals with chiral magnets.

To compare skyrmions with merons, we see that these structures are similar from the perspective of local geometry near the defect core:  They both have the same double-twist structure in the orientational order.  However, they are quite different from the perspective of global topology:  Around a skyrmion, the orientational order goes to the same vertical orientation everywhere.  Hence, it is possible to pack many skyrmions together with uniform regions in between.  The whole lattice of skyrmions is nonsingular, with approximately uniform magnitude of orientational order (uniform eigenvalues of $\bm{Q}$ for a liquid crystal, uniform $|\bm{M}|$ for a magnet).  By contrast, around a meron, the orientational order goes to a horizontal orientation, and it covers the full range of all possible horizontal orientations.  Hence, it is not possible to pack many merons together with uniform regions in between.  Rather, there must be singularities in the orientational order between the merons.  Hence, a lattice of merons can only form if the energetic cost of forming these singularities is not too great.

Because skyrmions are surrounded by uniform vertical orientational order, they can be regarded as local excitations.  Hence, skyrmions move and interact as effective particles, with only a short-range potential between them~\cite{Lin2013}.  Conversely, because merons are surrounded by the full range of nonuniform horizontal orientational order, they are more complex nonlocal excitations, which have long-range logarithmic interactions, and which must be accompanied by other defects,  This distinction in locality has been pointed out in the magnetic context~\cite{Nagaosa2013}, and it applies also in the liquid crystal context.

To compare chiral liquid crystals with chiral magnets, we note that these materials are similar from the perspective of topology:  They both can form skyrmions and merons.  However, chiral liquid crystals and chiral magnets are quite different from the perspective of energetics:  In chiral magnets, it is straightforward to stabilize skyrmions by applying a magnetic field, which couples linearly to $\bm{M}$ and stabilizes the orientation outside the skyrmions, in contrast with the orientation inside the skyrmions.  Hence, a lattice of skyrmions becomes the ground state for an appropriate choice of field and anisotropy.  By contrast, in chiral liquid crystals we have a tensor order parameter $\bm{Q}$, so there is no field that can distinguish between orientational order upward or downward; there is only a quadratic easy-axis or easy-plane anisotropy.  As a result, the specific model studied here does not have skyrmions as a ground state; it only has skyrmions as metastable defects.  To be sure, variations on this liquid-crystal model (perhaps with anisotropy arising from surface anchoring) might have skyrmions as a ground state, as in Ref.~\cite{Afghah2017}.  Even so, they are stabilized by a fairly delicate balance of free energies, not by the simple field as in the magnetic case.  Thus, we would state that the vector order parameter of magnets tends to favor skyrmions, while the tensor order parameter of liquid crystals tends to disfavor skyrmions.

In chiral liquid crystals, the formation of merons in a hexagonal lattice requires singularities of topological charge $-1/2$ between the merons.  In  ``high-chirality''  liquid crystal materials, the energetic cost of these singularities is not too large compared with the energetic benefit of the double-twist regions.  Hence, it is straightforward to stabilize meron lattices in liquid crystals.  Such lattices are called blue phases, and they have been studied extensively for many years.  In 3D liquid crystals, blue phases normally have a more complex cubic structure rather than the 2D lattice considered here, but still the same principles apply.  By contrast, in chiral magnets, the formation of merons in a square lattice requires singularities of the larger topological charge $-1$ between the merons.  It is theoretically possible for this structure to be the ground state, but it is difficult to find parameters where the energetic cost of the singularities is less than the energetic benefit of the double-twist regions.  Thus, we would state that the tensor order parameter of liquid crystals tends to favor merons, while the vector order parameter of magnets tends to disfavor merons.

\acknowledgments

We would like to thank S. Afghah and R. L. B. Selinger for helpful discussions.  This work was supported in part by National Science Foundation Grant No.~DMR-1409658 and in part by the U.S. Department of Energy.

\bibliography{skyrmions_paper_7}

\end{document}